\documentclass[aps, prx, onecolumn, superscriptaddress, longbibliography]{revtex4-2}

\usepackage{url}
\usepackage{braket}
\usepackage{graphicx}
\usepackage{multirow}
\usepackage{amsthm}
\usepackage{soul}
\usepackage{verbatim}
\usepackage{bm}
\usepackage{outlines}
\usepackage[exponent-product=\cdot]{siunitx}
\usepackage[space]{grffile}
\usepackage[dvipsnames]{xcolor}
\usepackage{xr}
\usepackage{color}
\usepackage{tikz}
\usetikzlibrary{calc, positioning}
\usepackage{mathtools}
\usepackage{algorithm}
\usepackage{algorithmic}
\usepackage{amsmath}
\usepackage{amssymb}
\usepackage{comment}
\usepackage[english]{babel}
\usepackage{subfigure}
\usepackage{booktabs}

\newcommand\myshade{85}
\colorlet{mylinkcolor}{violet}
\colorlet{mycitecolor}{YellowOrange}
\colorlet{myurlcolor}{Aquamarine}
\usepackage{hyperref}
\hypersetup{
	linkcolor  = mylinkcolor!\myshade!black,
	citecolor  = mycitecolor!\myshade!black,
	urlcolor   = myurlcolor!\myshade!black,
	colorlinks = true,
}
\usepackage[capitalize]{cleveref}

\graphicspath{{./images/}}

\begin{document}
	
    \title{Real-Time Decoding for Fault-Tolerant Quantum Computing: \\ Progress, Challenges and Outlook}
    
    \newcommand{\Qblox}{\affiliation{Qblox, Elektronicaweg 10, 2628 XG Delft, The Netherlands}}
    \newcommand{\UniMelbourneB}{\affiliation{School of Physics, The University of Melbourne, Parkville, 3010, Australia}}
    \newcommand{\CSIRO}{\affiliation{Data61, CSIRO, 3168, Clayton Australia}}
    \newcommand{\AWS}{\affiliation{AWS Center for Quantum Computing, Pasadena, CA 91125, USA}}
    \newcommand{\IQIM}{\affiliation{IQIM, California Institute of Technology, Pasadena, CA 91125, USA}}
    \newcommand{\QutechQCE}{\affiliation{QuTech and Department of Quantum and Computer Engineering, Delft University of Technology, 2600 GA Delft, The Netherlands}}
    \newcommand{\Riverlane}{\affiliation{Riverlane Ltd, Cambridge, United Kingdom}}
    \newcommand{\UniTokyo}{\affiliation{Graduate School of Information Science and Technology, The University of Tokyo, Tokyo, Japan}}
    \newcommand{\TUM}{\affiliation{Department of Computer Engineering, Technical University of Munich, Garching, Germany}}
    
    \author{F.~Battistel}\Qblox
    \author{C.~Chamberland}\AWS\IQIM
    \author{K.~Johar}\Riverlane
    \author{R.W.J.~Overwater}\QutechQCE
    \author{F.~Sebastiano}\QutechQCE
    \author{L.~Skoric}\Riverlane
    \author{Y.~Ueno}\UniTokyo\TUM
    \author{M.~Usman}\UniMelbourneB\CSIRO

    \date{\today}
    
    \begin{abstract}
        Quantum computing is poised to solve practically useful problems which are computationally intractable for classical supercomputers.
        However, the current generation of quantum computers are limited by errors that may only partially be mitigated by developing higher-quality qubits.
        Quantum error correction~(QEC) will thus be necessary to ensure fault tolerance.
        QEC protects the logical information by cyclically measuring syndrome information about the errors.
        An essential part of QEC is the decoder, which uses the syndrome to compute the likely effect of the errors on the logical degrees of freedom and provide a tentative correction.
        The decoder must be accurate, fast enough to keep pace with the QEC~cycle (e.g., on a microsecond timescale for superconducting qubits) and with hard real-time system integration to support logical operations. 
        As such, real-time decoding is essential to realize fault-tolerant quantum computing and to achieve quantum advantage. 
        In this work, we highlight some of the key challenges facing the implementation of real-time decoders while providing a succinct summary of the progress to-date. 
        Furthermore, we lay out our perspective for the future development and provide a possible roadmap for the field of real-time decoding in the next few years. 
        As the quantum hardware is anticipated to scale up, this perspective article will provide a guidance for researchers, focusing on the most pressing issues in real-time decoding and facilitating the development of solutions across quantum and computer science.
    \end{abstract}
    
    \maketitle

    \section{Introduction}
    \label{sec:intro}

            \begin{figure*}
                \centering
                \includegraphics[width=1\columnwidth]{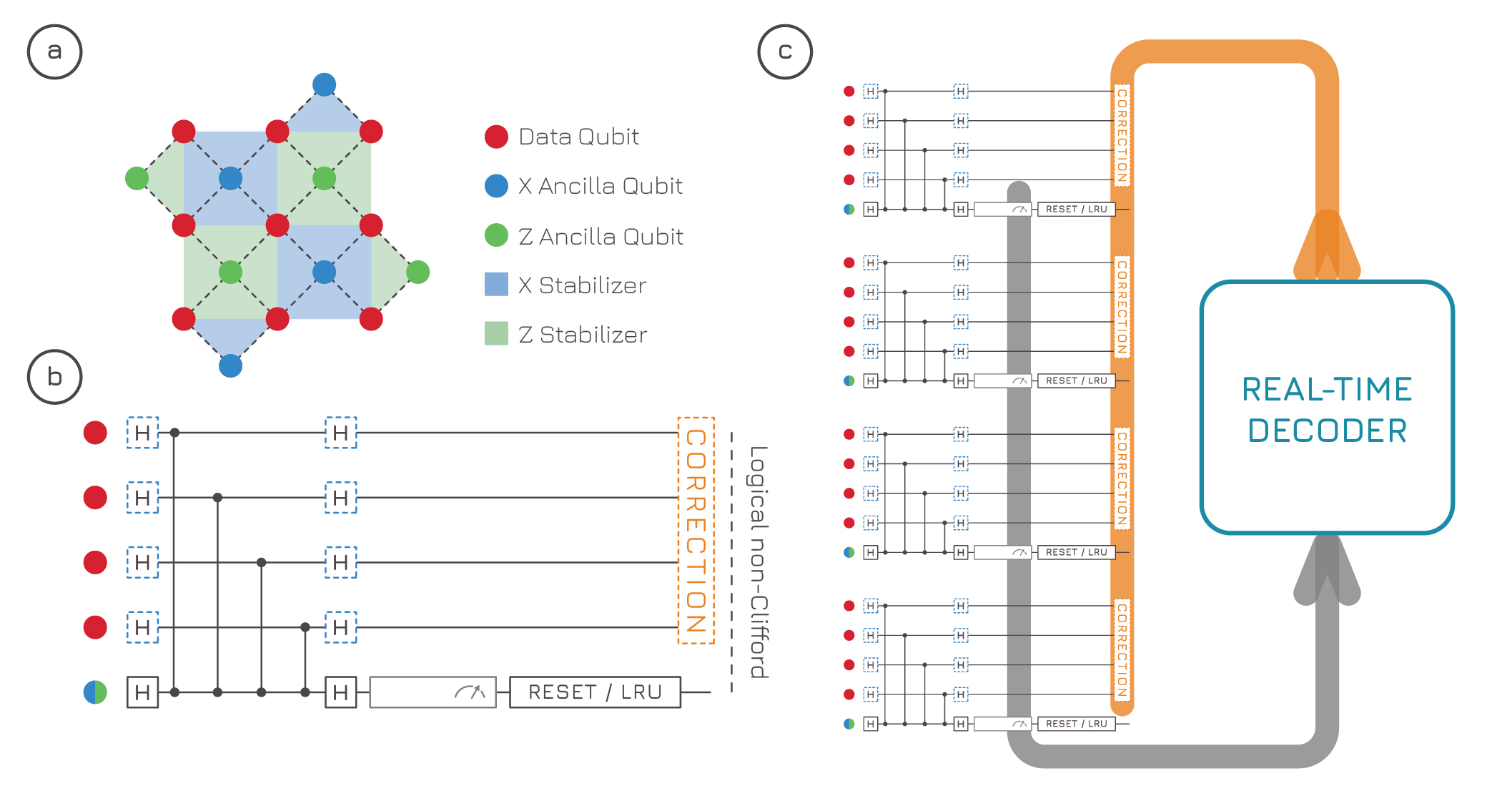}
                \caption{
                Quantum error correction and the role of real-time decoding.
                \textbf{(a)}~Distance-3 surface code.
                The logical information is encoded in the data qubits (pink).
                Ancilla qubits (blue, green) are used to measure error syndromes at every QEC~cycle.
                \textbf{(b)}~QEC~cycle for one stabilizer.
                The ancilla qubit performs CZs, is measured and then reset and/or a leakage-reduction unit are applied.
                The time from the last considered measurement to an impending logical non-Clifford gate is the available time for real-time decoding.
                \textbf{(c)}~The measurement outcomes from all stabilizers (usually including a few previous rounds) are fed to the real-time decoder, which updates the Pauli frame and outputs a correction.
                }
                \label{fig:decoding_realtime_intro}
            \end{figure*}

       Quantum computing has the potential to offer a revolutionary impact on both fundamental and applied sciences, leading to the solution of many computationally intensive problems which are currently intractable on classical supercomputers. 
       Recent advances in both quantum hardware and software fronts have brought the practical realisation of quantum computing possibly within a decade timeline. 
       Early demonstrations of quantum advantage~\cite{arute2019quantum, Zhong20} are salient examples indicating the dawn of a quantum revolution. 
       However, there remain serious challenges which must be resolved before a practical quantum advantage can be realised. 
       Among these, the effect of noise is one of the leading issues for the current generation of quantum devices, also known as Noisy Intermediate-Scale Quantum (NISQ) devices~\cite{Preskill18}. 
       Even with the anticipation of error rates tracking below~1\% or less, their cumulative impact on the execution of quantum algorithms requiring deep quantum circuits will be detrimental. 
       Therefore, the development and implementation of sophisticated quantum error correction~(QEC) schemes is imperative to achieve practical quantum advantage.     
       
        QEC is coming to the forefront as one of the key research areas towards the vision of fault-tolerant quantum computation~(FTQC) on faulty qubits in the near to long-term hardware systems. 
        To achieve FTQC, a multidisciplinary effort will be required with expertise ranging from quantum hardware, control systems and classical hardware design to error-correction protocols and computer science. 
        Fundamentally, QEC requires two schemes as illustrated in~\cref{fig:decoding_realtime_intro}. 
        Firstly, a code specifies an encoding in which physical degrees of freedom are combined to redundantly represent ``logical'' qubit states~\cite{Gottesman97,dennis2002Topological}.
        Information about the effects of errors on the logical qubits is obtained by cyclically measuring ancilla qubits yielding an error syndrome.
        Secondly, a classical decoder coprocessor uses a decoding algorithm to determine an appropriate correction to the logical states based on the syndrome data.
        If the underlying physical noise rate is below a certain threshold, the fault-tolerance threshold theorem states that QEC can lead to an arbitrary suppression of logical errors using a poly-logarithmic overhead in the number of qubits and computation time~\cite{KITAEV20032}.
        For the surface code the error rate has to be below a threshold of around~1\%, even though it should track below~0.1\% to make the overhead practically manageable.
        The correction established by the decoder is needed only for the next-in-line logical non-Clifford gate~\cite{Riesebos17} (for example to determine whether a logical~$S$ correction has to be applied during magic-state injection), otherwise it can be tracked by the decoder in the Pauli frame (or leakage frame~\cite{Suchara14}) until needed.
        A key parameter for the QEC~decoders is the delay introduced by the decoder itself.
        In fact, as the next logical non-Clifford gate requires a decoder-informed correction, the decoder needs to process the syndrome data at the same rate as it is received or close to it, i.e., in real time, to avoid an exponential slowdown of the computation, known as the \textit{backlog problem}~\cite{Terhal15}. 
        The incoming data rate varies for different qubit technologies and therefore the expected throughput requirements for real-time decoders range from~$\lesssim \mathcal{O}(1)~\mu$s for a surface code using superconducting transmon qubits~\cite{VanDijk19,Jeffrey14} 
        to~$\lesssim \mathcal{O}(1)$~ms for silicon spin-qubits~\cite{Gorman16,takeda2022Quantum,Barthel10,Vandersypen17} and even beyond $\mathcal{O}(100)$~ms for ion traps~\cite{Ryan-Anderson21}.
        Since the widespread transmons pose the tightest constraints, the value of~$1~\mu$s is often used as the benchmark for real-time decoding.
        However, this could be relaxed to a few~$\mu$s with the use of parallelization~\cite{skoric2022Parallel,tan2022Scalable}
        or trading time for space with the use of Auto-T gadgets (see Fig.17b in Ref.~\cite{litinski2019Game}), where the use of an extra ancilla per $T$~gate allows to postpone the moment when the output of the decoder is needed.
        Ultimately, the choice of the decoder architecture and its hardware will heavily depend on the target qubit technology.
        Identifying the optimum solution for each technology is an outstanding research question.    
        
\begin{figure*}
                \centering
                \includegraphics[width=0.8\columnwidth]{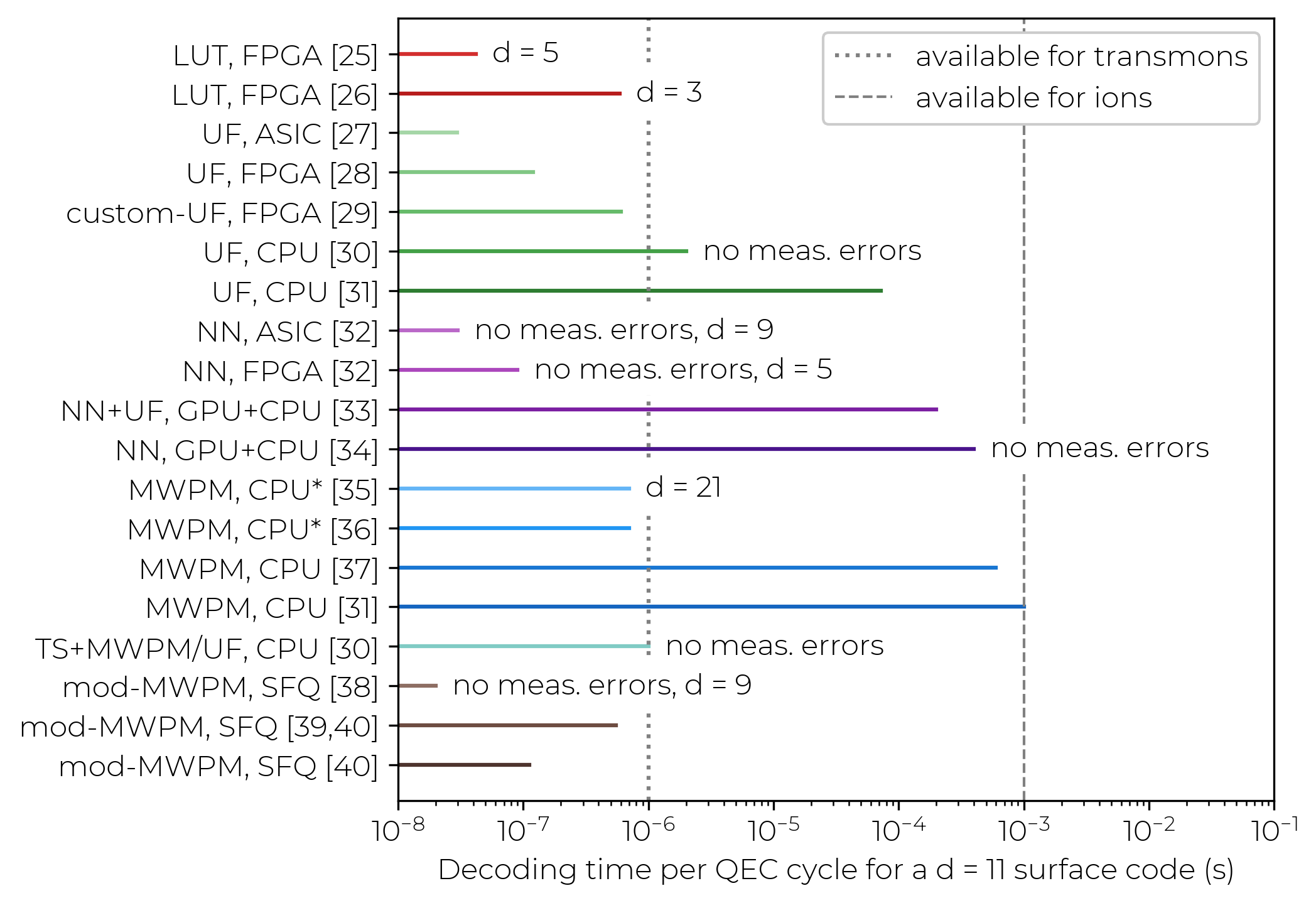}
                \caption{
                Comparison of Decoders: Summary of decoders for a distance-11 surface code (unless otherwise specified)~\cite{Das21,Riste19,Das22,Liyanage23,Riverlane22,Delfosse20,Huang20,Overwater22,meinerz2021scalable,Gicev21,fusion-blossom,pymatchingv2paper,fowler2012towards,Holmes20,Ueno21,ueno2022qulatis}.
                For the cases that do not account for measurement errors, the decoding time is given for a $d\times d\times 1$~block, whereas for those that do is given as $1/d$ of the time to decode a $d\times d\times d$~block.
                We do not divide by~$d$ only for the LUT~decoders given that the access time of a look-up table is fairly constant.
                Distance-11 is chosen since it represents a target that is challenging to implement but that at the same time might be attainable within a few years.
                We include the papers that report the decoding time.
                If no data point was available for distance-11, we extrapolate it from a fit of the available data if possible (if not, we specify the distance~$d$).
                Decoder types are marked by similar color shades: look-up table~(LUT, red), union-find~(UF, green), neural network~(NN, purple), minimum-weight perfect-matching~(MWPM, blue), two-stage~(TS, cyan), modified MWPM (brown).
                The labels on the left report the decoder type and the classical computational resources used in the given paper.
                SFQ refers to single-flux-quantum logic.
                The star near CPU refers to the time of the bare computation.
                By choosing a uniform distance we enable a comparison between decoders, even though the physical error rates vary.
                Indeed, the speed of especially UF and MWPM depends on the physical error rate, since more detection events require longer to be matched.
                However, the physical error rates are fairly uniform, being 5-10\% in the case without measurement errors, and 0.1-0.2\% if measurement errors are considered (phenomenological noise).
                Vertical dotted or dashed lines represent estimates of the available time to compute a correction and feed it back to the qubits on time.
                For superconducting transmon qubits the estimate~(1~$\mu\mathrm{s}$) is based on Refs.~\cite{Krinner22,Marques21}. 
                For trapped ions we take 1000~times that~(i.e., 1~ms) since the clock speed of ions is much lower than transmons, even though the QEC-cycle time of 200~ms reported in Ref.~\cite{Ryan-Anderson21} suggests that the bounds for ions might be even much looser.}
                \label{fig:comparison_decoders}
            \end{figure*}

        So far, the best known experimental demonstrations of real-time decoding have been with ion-trap based qubits~\cite{Ryan-Anderson21,egan2021Faulttolerant}. 
        It should be noted that these demonstrations in ion traps have not been based on algorithmic decoding but on a precomputed look-up table, which was sufficient given the small scale of the experiment. 
        On the contrary, experimental demonstrations on superconducting transmon qubits have been based only on offline decoding of errors for a quantum memory~\cite{Acharya22,Marques21,Krinner22,Sundaresan22}.
        However, a few proposals have shown potential to allow for real-time decoding in the near term~\cite{Das21,Riste19} and up to a fairly-large distance~\cite{Das22,Liyanage23,Riverlane22,Delfosse20,Huang20,Overwater22,meinerz2021scalable,Gicev21,fusion-blossom,pymatchingv2paper,fowler2012towards,Holmes20,Ueno21,ueno2022qulatis} (see~\cref{fig:comparison_decoders} for an overview).
        Despite these promising results, the experimental implementation of an entirely functional real-time decoding framework fully compatible with a scalable quantum processor still requires substantial development in the next few years. 

            \begin{figure*}
                \centering
                \includegraphics[width=1\columnwidth]{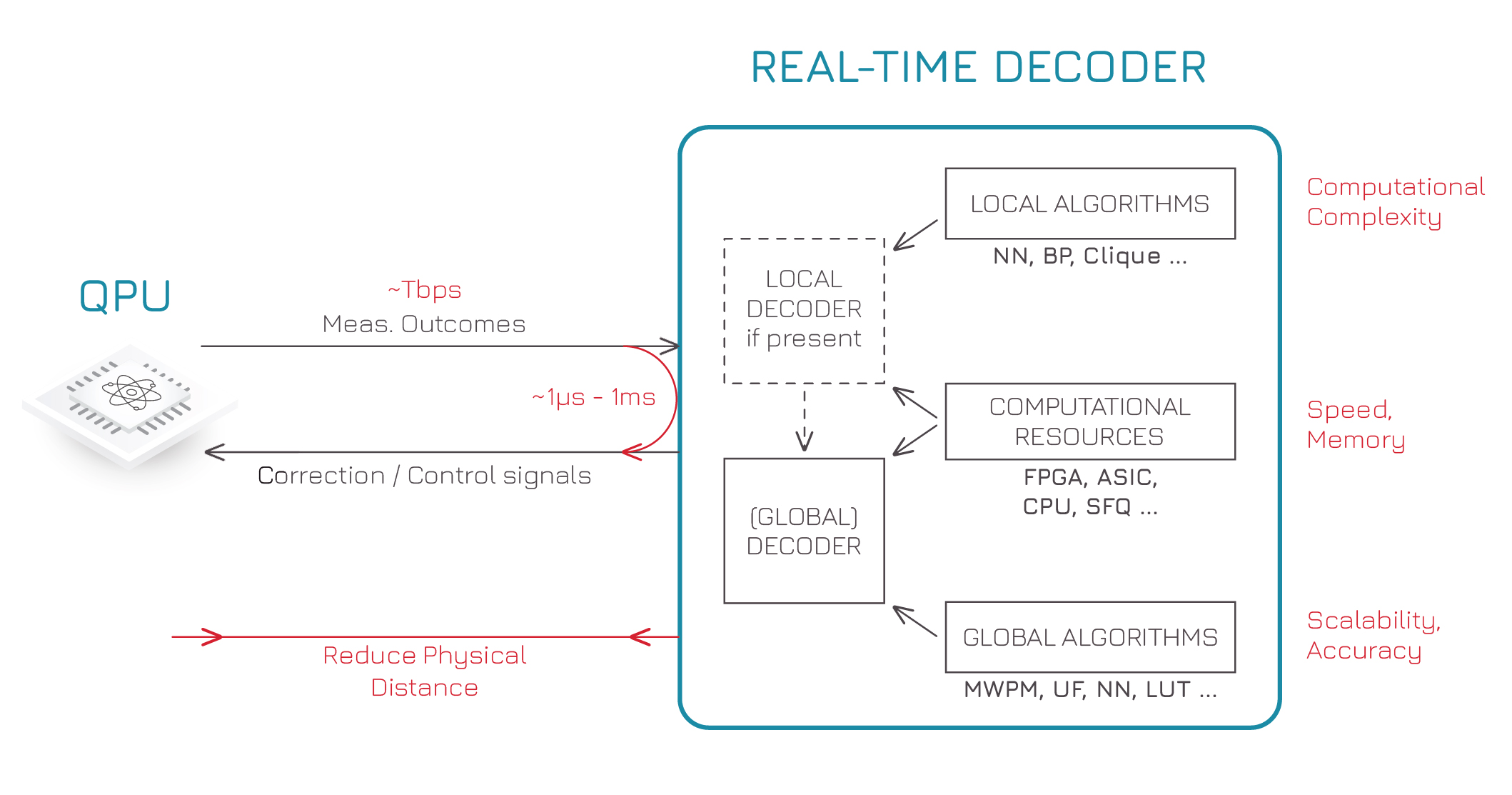}
                \caption{
                Real-time decoding framework and associated challenges (red).
                Measurement outcomes from the ancilla qubits in the QPU are passed to a local decoder, if present, otherwise directly to the global decoder.
                The decoder updates its estimate of the Pauli frame of the qubits (as well as the leakage frame possibly).
                Physical correction pulses are sent to the QPU when requested because of an impending logical non-Clifford gate (e.g.~to apply or not a logical~$S$ correction during magic-state injection).
                A few possible decoding algorithms and computational resources are listed.
                }
                \label{fig:decoding_realtime_blockdiagram}
            \end{figure*}

        In this perspective article we provide a succinct summary of the current status with respect to each challenge and a roadmap for the future development, which will serve as a blueprint to guide quantum and computer-science researchers working in the areas of QEC and real-time decoders. 
        While the real-time decoding field is still in its infancy, we believe that its steady growth makes it timely for a review and perspective to boost a further expansion of the real-time decoding community.
        \Cref{fig:decoding_realtime_blockdiagram} plots the flowchart diagram of a real-time decoder framework, highlighting key challenges associated with each component. 
        The decoding algorithm must be scalable and have a limited computational complexity, 
        while the underlying classical computational resources need to have enough speed and memory.
        Not only the decoding algorithm itself has to run fast, but the deployment of measurement outcomes to the decoding module and feedback in the quantum control stack must keep a fast pace, even though this requirement could be mitigated using selective teleportation~\cite{fowler2013timeoptimal} at the cost of slowing down the quantum computation and of a higher error rate per logical gate unless an increased code distance is employed.
        Hence, the communication latency, bandwidth and location of the decoding module are crucial.
        Furthermore, the decoder must be applicable not just to quantum memories but also to logical operations such as lattice surgery. 
        We first discuss the requirements for real-time decoding and hard system integration in~\cref{sec:requirements}.
        We then cover the challenges and perspectives related to the decoding algorithms in~\cref{sec:decoding_algorithms}, and to the computational resources, their location and latency in~\cref{sec:hardware_cost}.
        We conclude with an outlook on the next few years in~\cref{sec:conclusion}.

    \section{Requirements for (hard) real-time decoding}
    \label{sec:requirements}

        Different platforms are going to require custom-made decoding algorithms to address individual requirements such as decoding rate, noise profiles, and device topologies, as well as cost. For superconducting and photonic qubits, a high operational rate implies that the decoding throughput and latency are paramount~\cite{Acharya22,bartolucci2021Fusionbased}.
        Throughput (defined as the time it takes the decoder coprocessor to perform the decoding once it receives the measurement outcomes) is even more important than latency (defined as the time it takes to send the measurement outcomes to the decoder coprocessor)~\cite{Chamberland22,fowler2013timeoptimal}. 
        In recent superconducting-qubit experiments, QEC~rounds were performed every $\sim 1~\mu$s~\cite{Acharya22,Marques21,Krinner22,Sundaresan22}, leading to an estimate that a utility-scale quantum computer would generate a few tens of Mbps of syndrome data per logical qubit, for a total of many Tbps~\cite{bacondave2022Software}. For each logical qubit, the decoder needs to process that data at the acquisition rate or close to it 
        to avoid an exponential slowdown due to an ever-growing data backlog.
        
        In the case of ion-trap and neutral-atom quantum computing, the QEC-cycle times are usually in excess of many milliseconds~\cite{Ryan-Anderson21, egan2021Faulttolerant, webber2022impact, henriet2020Quantum}, leaving ample time for real-time decoding. We note that relatively slow operations in these platforms are compensated by high connectivity and long coherence times. These features allow to choose across a wide landscape of QEC~codes and their decoders, beyond the surface code.
        For example, color codes have been studied for this purpose due to lower qubit overheads and the ease of applying fault-tolerant operations~\cite{landahl2011Faulttolerant, Ryan-Anderson21}.
        While polynomially-efficient decoders for color codes have been proposed~\cite{sarvepalli2012Efficient, kubica2019Efficient}, further work is required in order to improve performance in realistic conditions in the presence of circuit-level noise~\cite{chamberland2020Triangular}.
        Furthermore, the increased connectivity can be exploited to implement quantum low-density parity-check codes~(LDPC) which outperform surface and color codes in terms of protection against errors and the ratio between logical and physical qubits~\cite{breuckmann2021Quantum, panteleev2022Asymptotically, gottesman2014FaultTolerant}.
        However, quantum LDPC~codes may require lower physical error rates due to lower thresholds than surface codes.
        While classical LDPC~codes have fast decoders, these do not generalize well to their quantum counterparts.
        With a few algorithms starting to investigate this space~\cite{roffe2020Decoding, gu2022efficient}, we expect that over the coming years this gap is going to be closed, bringing quantum LDPC~codes closer to practical applications.
        
        The first experiments with error-corrected silicon~\cite{takeda2022Quantum} and solid-state~\cite{abobeih2022Faulttolerant} qubits are also beginning to take place.
        We expect that the encoding and decoding requirements for these platforms are going to be defined over the coming years as the systems mature.

    \subsection{Hard versus soft real-time}
    \label{sec:hardsoftRT}

        The feedback loop of altering the direction of the program depending on the measured error syndromes can only be possible if the decoder is active and running during the execution of the quantum program.
        In addition to decoding being real-time, it is also important to establish the subtlety of \emph{hard} versus \emph{soft} real-time.
        In a hard real-time system~\cite{Burns91}, each computational piece must complete within a deterministic and precise amount of time.
        The sequencing of the individual events must be time-bounded and it is often statically determined by performing Worst-Case Execution Times analysis (WCETs)~\cite{Wilhelm08}.
        On the contrary, soft real-time systems are those that can tolerate some variation in how long the program takes to execute.
        Performing WCET~analysis is generally extremely tedious, complex and highly sub-optimal for systems that have a probabilistic behaviour.
        Even if the analysis can be performed, allocating fixed time to individual tasks and scheduling them statically could lead to overallocation and therefore lead to performance degradation.
        In the case of quantum computing, the lost time will also affect qubit coherence.

        A fault-tolerant quantum stack will likely need every piece to precisely execute within a fixed timing budget. These pieces range from the pulses that are generated to control the QPU, to repeatedly reading out intermediate states and correcting the direction of future pulses based on the outcome of decoding errors, to magic-state distillation~\cite{prabhu2022New}. As such, a fault-tolerant quantum computer seems to constitute a hard real-time system, even though some softness may have to be allowed for a highly non-deterministic process such as magic-state distillation, while QEC~cycles are repeated in a hard real-time fashion.
        Hard real-time decoding requires tighter system integration to remove any unnecessary communication overhead and latencies. This sometimes implies that specific customizations will need to be made to every part of the stack, which increases development time and costs. Furthermore, tight integration will have to be conjugated with scalability of the system.
        As explained above, the determination of an optimal schedule for a hard real-time system is tedious and complex.
        Therefore, high-level execution models for various parts of the stack will need to be put together to determine its schedule and time allocation.
        The use of WCET~tools and methodologies will help in reducing or removing any slack in the execution and help improve the performance of the quantum program while reducing the chances of decohering.

    \section{Decoding algorithms}
    \label{sec:decoding_algorithms}

        In principle, every code can be decoded with a maximum-likelihood decoder by assigning a probability to each set of errors that would generate the given syndrome.
        These can then be separated into equivalence classes based on their effect on the logical degrees of freedom.
        The class with the greater total likelihood is then taken as the decoding result.
        While this provides the optimal logical fidelity, in general it has been shown to be a $\sharp P$-complete problem~\cite{iyer2013Hardness}.
        Tensor network decoders reduce the problem of optimal decoding to contractions of tensor networks that can be made more efficient by arbitrarily-precise approximate contractions~\cite{ferris2014Tensor, chubb2021General,bravyi2014Efficient}, but nevertheless require substantial computational resources. 
        Therefore, as much as designing qubit-efficient and robust codes, the challenge of QEC consists of finding less-accurate decoders that achieve the required trade-offs between logical fidelity and classical computing requirements.

        One general solution is to pre-compute the corrections for a given syndrome in a look-up table, such as LILLIPUT~\cite{Das21}.
        While extremely fast and relatively easy to implement, the memory requirements of look-up tables scale exponentially, making them impractical but for the smallest demonstrations.
        A more scalable approach is to design fast algorithms that are tailored to a smaller family of codes and that can be efficiently implemented. 
        For the widespread topological codes like the surface code, these include minimum-weight perfect matching (MWPM)~\cite{dennis2002Topological,Edmonds73}, union-find (UF)~\cite{Delfosse21,Huang20}, renormalization group~\cite{duclos-cianci2010Fast} and belief propagation~\cite{Criger18,higgott2022Fragile,roffe2020Decoding} decoders (the latter cannot be used as a standalone decoder but can supplement e.g.~MWPM or UF for improved performance).
        While significant engineering effort is required to realize these implementations, recent improvements in parallelization of real-time decoders~\cite{skoric2022Parallel, tan2022Scalable} warrant optimism that scalable decoding hardware is going to be demonstrated in the forthcoming years.

        To date, all hardware demonstrations of quantum error correction have fallen short of unambiguously achieving logical lifetimes better than physical lifetimes, though several come close~\cite{Ryan-Anderson21,Krinner22,egan2021Faulttolerant,abobeih2022Faulttolerant,zhao2022Realization,Sundaresan22,Acharya22}.
        As near-term hardware is likely going to suffer considerable error rates, to unequivocally demonstrate this next big technological milestone, accurate decoders that are closely tuned to the physical noise are going to be necessary.
        Maintaining high fidelity in the presence of correlated errors~\cite{paler2022Pipelined},  leakage~\cite{Marques21,Varbanov20,Battistel21}, erasure~\cite{delfosse2020LinearTime}, device inhomogeneities~\cite{Krinner22}, and even rare high-energy events~\cite{McEwen21} necessitates modifications to standard decoders.
        This also means that any decoding hardware is going to have to be adaptable to the characteristics of a particular device, and tuned as the noise profile changes with time.

    \subsection{Minimum weight perfect matching (MWPM) and union find (UF)}
    \label{sec:MWPMandUF}

        Topological surface codes have been the focus of QEC~research as a promising architecture for scalable quantum computation~\cite{fowler2012Surface, dennis2002Topological} due to a low connectivity requirement and high tolerance to errors.
        Moreover, there is an abundance of literature on how to efficiently perform fault-tolerant quantum computation in a surface-code architecture~\cite{litinski2019Game, chamberland2021Universal}.
        Local fault mechanisms in surface codes can be decomposed into errors that trigger either a pair of changes or a single change in the syndrome, referred to as defects or detection events.
        This makes them suitable for decoding with a minimum-weight perfect matching (MWPM) algorithm~\cite{dennis2002Topological,fowler2014Minimum}.
        MWPM finds the smallest error consistent with the syndrome in polynomial time using the blossom algorithm~\cite{Edmonds73,kolmogorov2009Blossoma}.
        While MWPM does not take into account the degeneracy present in QEC~codes~\cite{beverland2019role}, it nevertheless achieves good performance even against circuit-level noise.
        Fast implementations of MWPM have been recently released, namely fusion-blossom~\cite{fusion-blossom} and sparse-blossom (PyMatching~v2)~\cite{higgott2021PyMatching,pymatchingv2paper,pymatchingv2}.
        Both exploit the sparse decoding graph to solve MWPM more efficiently.
        Sparse-blossom uses further optimization strategies to grow alternating trees and fusion-blossom allows parallelization over many spacetime chunks that are then fused together.
        The two approaches may be combined together to achieve further speedup~\cite{pymatchingv2}.

        Another algorithm is union-find~(UF), which has close-to-linear runtime~\cite{Delfosse21, Huang20}, a distributed FPGA~realization~\cite{Liyanage23} and a proposed micro-architecture~\cite{Das22}.
        Furthermore, the three major steps in the algorithm could be handled by three different hardware units, some of which may be faster than the others.
        In a quantum computer with many logical qubits, a proposal~\cite{Das22} to be demonstrated in practice is to have multiple copies of the slower units but share the faster units across many decoders to reduce hardware overhead.
        This approach could also be applied to other types of decoders.

    \subsection{Neural-network decoders}
    \label{sec:neural_networks}

        Besides being fast, a decoder must also be accurate, scalable with respect to the number of physical qubits, able to tackle complex noise models and compatible with lattice surgery.
        In those respects, neural-network decoders are promising candidates for real-time decoding, thanks to their constant inference time, the inherent ability to learn any error model, the scalability to large code distances~\cite{Ni20, Gicev21, Chamberland22} and compatibility with lattice surgery~\cite{Gicev21,Ueno22,Chamberland22}.
        Nevertheless, several challenges must be overcome before seeing neural-network hardware decoders in a practical quantum  computer, such as finding the optimal neural-network architecture able to address complex error models, and quantifying the trade-offs between the hardware costs and the decoder performance.
        Furthermore, as of yet, the performance against time-like errors during lattice surgery (see~\cref{sec:lattice_surgery}) when using pre-decoders (see~\cref{sec:two_stage}), such as neural networks, has not been analyzed.
        The adaptability to change in surface-code patch shapes through time has to be demonstrated as well.

        The training strategy is also a crucial choice for neural-network  decoders.
        Supervised learning is the straightforward choice when the training data is synthetically generated using a software-implemented error model as both the decoder input and its expected output for the full training dataset are known, e.g.~\cite{Overwater22,meinerz2021scalable,Gicev21,Ueno22,Ni20}. 
        Nevertheless, reinforcement learning has also shown promise~\cite{Andreasson2019quantumerror, DomingoColomer2020}, and is particularly interesting when training on data from real quantum hardware, for which only the syndrome is known but the errors on the data qubit are not. 
        Furthermore, reinforcement learning can be applied to automatically discover new QEC~schemes and their corresponding decoders by exploiting the feature of a specific quantum platform to gain efficiency\cite{Fosel2018, Nautrup2019optimizingquantum, Zeng2022arXiv} and also leading to a recent real-time implementation for superconducting qubits~\cite{Sivak2023}.

        Hardware-based decoder implementations have been proposed as alternative to software decoders to keep the (expected) decoding times below the $1~\mu$s~target, e.g., employing FPGAs~\cite{Overwater22,Das21, Liyanage23, Chamberland22,Riste19,Riverlane22}, ASICs~\cite{Overwater22,Das22} or Single-Flux Quantum logic (SFQ)~\cite{Holmes20,Ueno21,ueno2022qulatis,Ueno22}.
        In the case of neural-network-based decoders, implementing them in hardware can exploit the advances in the field of hardware accelerators for neural networks for classical applications, such as image and voice recognition.
        These hardware-accelerator ASICs have been recently the subject of a fast-evolving research direction, where the main emphasis lies on improving the underlying hardware fabric while also exploiting the sparsity of the network~\cite{Murmann20}, e.g., by co-optimizing the quantization in the digital signal representation and pruning~\cite{Han15,Esser19}.
        In terms of hardware costs, state-of-the-art fully-digital accelerators currently achieve power efficiency up to 1000~TOPS/W with area efficiency of 400~TOPS/mm$^2$ in integrated-circuit implementations~\cite{Knag20}.
        Since the main limitations with these architectures are the bandwidth and power bottleneck in fetching and transferring data between the memory and the computational unit, in-memory computing has been proposed for a better scaling in energy, bandwidth and delay~\cite{Verma19}.
        Additional improvement in efficiency for in-memory-computing neural networks can in principle be achieved by moving from fully-digital solutions to analog and mixed-signal ones, which already demonstrate up to 500~TOPS/W and are quickly closing the gap with the digital solutions~\cite{Yin21}.
        However, these analog solutions are still lagging behind in area efficiency, although non-CMOS technologies, such as flash- and memristor-based architectures, promise significant enhancements in that direction.
        Coming years will be crucial to demonstrate whether such progress can be applied also to neural-network decoders for QEC.

        On the way to 2025, quantum-computing systems are quickly growing to a scale where experiments with useful QEC~schemes can be performed.
        The hardware optimization will then become more relevant and several questions must be addressed in the next few years towards the goal of a real-time neural-network-based decoder.
        The best network architecture must be identified: for instance, while Convolutional Neural Networks (CNN) seem amenable to scalability and lattice surgery, which architecture can best address measurement errors: a 3D~CNN or a recurrent 2D~CNN?
        Following the examples set by state-of-the-art classical neural-network hardware, what are the trade-offs between decoding performance (accuracy, delay) and hardware costs (power, area)?
        Can neural-network decoders be efficiently adopted for emerging QEC~schemes?
        Which role can reinforcement learning play?
        The resulting tight constraints in area and power will push towards advanced techniques in neural-network hardware, such as analog computing and in-memory computing, possibly in a cryo-CMOS chip hosting qubits, interface electronics and QEC~decoders as well (see~\cref{sec:cryogenic} for more about this).

    \subsection{SFQ-based decoders}
    \label{sec:SFQ_decoders}

        Single Flux-Quantum~(SFQ) logic is a digital circuit composed of superconductor devices, and it has the potential for utilization in next-generation computers because of its ultra-fast and low-power performance compared to CMOS~\cite{Likharev91,Kirichenko11}.
        Information processing in SFQ circuits is performed with magnetic flux quanta stored in superconductor rings. The presence or absence of an SFQ in the ring represents a logical `1' or `0', respectively. 
        The pulse-driven nature of information processing in SFQ circuits enables both their fast switching~($\sim10^{-12}$~s) and low-energy consumption~($\sim10^{-19}$~J per switching).
        Hence, with this technology it is feasible to increase the device clock frequency to~$\mathcal{O}(10$-$100)$~GHz while keeping its power consumption low enough to operate in a cryogenic environment.
        On the downside, SFQ-based decoders need to be designed and developed by hand (called custom design).
        SPICE~simulations are then needed to prove that the output is as expected for given input conditions.
        To scale this development methodology to complex algorithms, SFQ~logic needs to be fully supported by commercial EDA~tools (Electronic Design Automation) that do not exist yet.
        
        Several SFQ-based peripherals have been proposed for fundamental operations of quantum computers, such as measurement and manipulation of superconducting qubits~\cite{Leonard18,Liebermann16,Johar22}.
        In addition, SFQ~circuits have been proposed for decoding QEC~codes~\cite{Holmes20,Ueno21,ueno2022qulatis,Ueno22}.
        However, memory-intensive decoding algorithms, such as MWPM or UF, are unsuitable for SFQ~circuits because a large amount of RAM is expensive in SFQ~implementations.
        Thus, memory-efficient decoding algorithms are required for SFQ~circuit execution.

        Ref.~\cite{Holmes20} proposed a new decoding algorithm for the surface code, named AQEC, and designed the first SFQ~chip implementing such algorithm, while the concept of using SFQ~circuits for QEC had already been proposed~\cite{tannu2017taming}.
        This decoder consists of multiple units corresponding to each data and ancilla qubit to detect and correct errors by propagating simple signals between the units in a distributed processing scheme.
        However, this decoder focuses only on correcting Pauli errors on data qubits and is incapable of addressing measurement errors.
        
        Ref.~\cite{Ueno21} proposed the QECOOL~algorithm, which is an extension of AQEC to deal with measurement errors, designing an SFQ-based chip which achieves even lower power consumption than the AQEC~decoder.
        Furthermore, QECOOL is an online-QEC or sliding-window decoder, which in general seems more appropriate to keep the decoding time bounded for realistic QEC, where new syndrome data comes in as input at every QEC~cycle.
        Ref.~\cite{ueno2022qulatis} proposed the QULATIS decoder by extending QECOOL to deal with lattice surgery (see~\cref{sec:lattice_surgery}) and Ref.~\cite{Ueno22} proposed the NEO-QEC decoder by combining a binarized neural-network-based SFQ~decoder in a two-stage process with QECOOL/QULATIS.

    \subsection{Two-stage decoders}
    \label{sec:two_stage}

        Many proposals~\cite{Delfosse20,Ueno22,Chamberland22,ravi2022better,meinerz2021scalable,Gicev21,Smith22,higgott2022Fragile} have been put forward where decoding occurs in two stages, with a first, computationally-simple local decoder and a second, more-complex global decoder.
        The first stage can either be a pre-processing stage whose outcome is passed anyway to the global decoder~\cite{higgott2022Fragile}, or it can try to correct all errors (if the pattern is ``simple'' enough) and call the global decoder only when it fails.
        In the latter case it can provide either the original~\cite{Delfosse20} or a pre-processed syndrome~\cite{Ueno22,Chamberland22,meinerz2021scalable,Gicev21,ravi2022better,Smith22} to the global decoder.

        The purpose of two-stage decoding can be to enhance the accuracy and/or speed of the global decoder, in which case the two stages can be co-located.
        For example, the techniques in Ref.~\cite{Chamberland22} improve the speed of MWPM for pure memory by~$10^6$ for circuit-level noise.
        However it is required that such local decoders are implemented in times less than~$\mathcal{O}(1)~\mu$s (for superconducting-qubit architectures) to be useful.
        In general, the main purpose of two-stage decoding is to reduce bandwidth on the cryostat~I/O, as well as the power consumption inside the cryostat, by placing the first stage at cryogenic temperature and the second stage at room temperature.
        In particular, the first-stage decoder must be computationally-inexpensive so that its power requirements are minimal.
        Provided low-enough physical error rates, low-weight and sparse error configurations will be the most common and thus can be handled by the local decoder alone, without having to often send signals to the top of the cryostat.
        For example, the Clique decoder~\cite{ravi2022better} has a hierarchical structure where a simple SFQ~decoder in the 4K~environment and a complex decoder in the room-temperature environment are combined.
        In this implementation, the SFQ~part decodes only the case where all weight-1 errors are isolated, and the more difficult error signatures are sent to the second-stage decoder.
        Furthermore, the SFQ~part is simple enough to achieve a much lower resource overhead than global SFQ-based decoders (see~\cref{tab:cryo_decoder_comparison}).

    \subsection{Lattice surgery}
    \label{sec:lattice_surgery}

        Several systems are now at the size where implementing multi-qubit logical operations in a fault-tolerant manner is becoming feasible.
        In surface codes, these are performed by merging and splitting code patches in a scheme known as lattice surgery~\cite{litinski2019Game, chamberland2021Universal,Chamberland22, ueno2022qulatis}.
        
        Lattice-surgery decoders take the full syndrome information into account over a given syndrome history window. 
        Corrections are applied in a manner identical to what would be done with a decoder used for pure memory with a few exceptions.
        First, to correctly address boundary effects obtained by merging and splitting surface code patches, logical representatives must be defined with care, since corrections can always be viewed as flipping the sign of a set of logical operators.
        As an example, when performing an $X\otimes X$ measurement via lattice surgery, ancilla qubits in the routing space region will initially be prepared in $Z$~basis eigenstates.
        As such, logical~$Z$ representatives for the surface code patches can be taken to have support on qubits in the routing space before merging the patches and after the split.
        Such re-definitions of the logical operators avoid ambiguities due to boundary effects.
        Second, certain lattice operations require measuring extended rectangles (such as weight-4 checks which are longer range or require more ancilla qubits) and potentially higher weight stabilizers (for instance if twists are used to measure~$Y$). 
        In such cases, the information from the additional ancilla qubits must be used with care to reconstruct the stabilizer measurement outcomes.
        With the above points, error correction can then proceed exactly the same way as what would be done with pure memory.
        For instance, if a MWPM decoder is used, matching would be performed over the syndrome history window, with each edge in the graph specifying which logical is flipped. 
        The same holds for any other graph-based decoder, such as union-find, as long as logical representatives are correctly specified through the full syndrome history of the computation.

        While performing lattice-surgery operations, the code patches are changing in size and shape as they move, merge, and split.
        This means that the decoding hardware is going to have to be reconfigurable on-the-fly, working in a low-latency feedback loop with the control system to implement complex fault-tolerant circuits.
        This process also often requires decoding excessively large codes during the merge stage.
        Therefore, the decoding will likely have to be parallelized across space, as well as time, in order to prevent a slowdown of the logical clock speed~\cite{skoric2022Parallel,tan2022Scalable}.

        High thresholds for space-like as well as time-like errors when performing lattice surgery still need to be demonstrated.
        Recall that time-like errors arise as an error in the parity of the multi-qubit Pauli measurement outcome when performing lattice surgery.
        Temporal encoding of lattice surgery~(TELS)~\cite{prabhu2022New,chamberland2021Universal} allows one to correct logical time-like failures which occur during lattice surgery using measurement outcomes from a redundant amount of lattice-surgery measurements.
        The decoding of TELS uses classical error correcting codes.
        This could be used to alleviate some of the decoding requirements, potentially in conjunction with dedicated control hardware for TELS.

    \section{Computational resources}
    \label{sec:hardware_cost}

    \subsection{CPUs, FPGAs, ASICs}
    \label{sec:CPU_FPGA_ASIC}

        The challenges faced around computational resources depend on the choice of the platform. Here we consider all three popular choices (CPUs, FPGAs, and ASICs) and highlight the relevant progress and challenges. 
        
        \textbf{CPUs}.
        The first challenge facing the execution of decoding algorithms on CPUs is that these are currently benchmarked on desktop/server class CPUs.
        The typical power consumed by a desktop computer is in the order of several watts.
        While it would be possible to use multiple threads on the machine to decode in parallel, there are significant challenges in scaling this to several thousands of logical qubits.
        The problem gets exacerbated if the control electronics moves into cryogenic environments, which have strict power budgets in the order of 1~mW per qubit at~4~K.
        On the contrary, there are CPUs that are already part of the room-temperature quantum control stack.
        These CPUs are on FPGA~boards which are used to control the sequences of pulses to the QPU.
        In most cases these CPUs are underutilized and therefore could be good candidates to use for real-time decoding.
        However, it must be noted that these CPUs are generally several generations old, are clocked at lower frequencies and therefore are at least an order of magnitude lower in performance compared to a typical laptop/server grade CPU.
        
        Another challenge that CPUs would face is around \emph{hard} real-time decoding (see~\cref{sec:hardsoftRT}).
        Typical usage of CPUs in the stack makes the system non-deterministic.
        CPUs are designed to run operating systems with memory-management units and address translation.
        They also have multiple levels of caching from level~1 to off-chip memory.
        The combination of using a non-real-time operating system such as Linux, memory management, memory hierarchies and out-of-order execution makes the system non-deterministic.
        This implies that it is almost impossible to determine the worst-case execution time of any computation (such as decoding).
        As explained in~\cref{sec:hardsoftRT}, for hard real-time systems having a bounded worst-case execution time is vital.
        The above problems can be worked around by using CPUs that are specifically designed for hard real-time execution, running bare metal without any operating system and avoiding the use of caches.
        However, this will negatively impact the performance of the system since real-time CPUs typically cater to a lower-performance category.
        Furthermore, in order to operate a tight loop between control systems and CPUs for decoding, it would be necessary to transport the data as close to the CPU~caches as possible.
        A good number of high-performance CPUs have accelerated coherency ports that allow pre-loading of caches, but there might be additional coherency-port overheads that lead to slowdown in the execution of the decoding algorithm.
        
        \textbf{FPGAs.}
        There have been several proposals of building decoders onto an FPGA~\cite{Das21,Chamberland22,Riste19,Overwater22,Liyanage23,Riverlane22}.
        The primary advantage of using an FPGA~decoder in the near term is to allow for its straightforward integration with existing control systems for low-level pulse control, which are also typically built on FPGAs.
        The vast majority of the control systems already have direct access to readout of qubits at the FPGA~level and therefore that information can be readily transported to the decoding FPGA via an appropriate bus protocol.
        This protocol has to be compatible with a distributed control-stack architecture and has to have extremely low latency, taking at most a couple hundreds of nanoseconds to not overburden the decoding budget.
        Another advantage is that FPGAs have relatively short development cycles for developing algorithms and are easily reconfigurable.
        It is also possible to get execution frequency in the order of 100s~MHz, even though this is still lower than with CPUs or ASICs in principle~\cite{Kuon09}. 
        Furthermore, FPGAs have recently demonstrated the execution of algorithms specifically designed for parallel hardware for a fairly large code distance~\cite{Liyanage23}.
        On the downside, high-end FPGAs used to control qubits are generally expensive, which would make it cost-prohibitive to scale to millions of physical qubits.
        
        \textbf{ASICs.}
        The landscape around the development of physical qubits and QEC is constantly evolving.
        It is expected to take several years before industry-wide consensus is reached on the precise requirements of a truly useful real-time decoder, including choices of decoding algorithms, QEC~codes, interfaces with the control system, etc.
        While the industry moves forward and active research converges on practical solutions, the use of ASICs would be premature due to the complex and cost-prohibitive nature of ASIC~development.
        This constitutes the primary obstacle for ASIC~adoption for decoding systems, temporarily outweighing the benefits of low per-part costs, drastically reduced power consumption, high execution frequencies in the order of 1-2~GHz and very tight integration for scaling.
        However, if the control electronics moves into the cryogenic environment, these advantages become natural arguments for complete decoder and control-electronics integration using ASICs, and a path forward to large-scale systems.

    \subsection{Location of the decoder module}
    \label{sec:location_QEC_module}

        The location of the decoder module and its subsequent latency and bandwidth will depend on the overall system stack.
        The system stack as it stands today is primarily built to improve the quality of the qubits, rapidly prototype improvements to the control systems and updates to the software stack.
        In the NISQ era, it is likely that the decoder will be integrated in different parts of the stack (see~\cref{sec:two_stage}).
        However, this picture is likely to be very different in the fault-tolerant era.
        A parallel is often drawn between classical and quantum computers: in fact, one of the highlights of classical computers has been the progress through Moore's law and the Very Large Scale Integration~(VLSI).
        The two together have contributed to significant improvements in the computational power, large reduction in size, energy requirements and most importantly bringing the cost down, making the technology accessible to almost everyone.
        A similar evolution is also envisaged for quantum computers and it is therefore natural that large-scale integration will enable quantum computers to scale.
        Given this paradigm shift, the ideal location for the decoder would be alongside the control electronics, integrated with very low latency in the order of 10s of nanoseconds with fast decoder execution and fast feedback into control for fault-tolerant logical operations.

    \subsection{Opportunities at cryogenic temperature}
    \label{sec:cryogenic}

        In many quantum-computing platforms qubits are operated in a cryogenic environment, such as inside of a cryostat, to reduce noise.
        Superconducting quantum computers require many high-frequency coaxial cables connecting qubits and peripherals for manipulation, measurement, and QEC to a room-temperature environment.
        The heat inflow and the occupied space by cabling in the cryostat seriously limit the scalability of superconducting quantum computers.
        The same problem applies to other types of solid-state qubits that operate in a cryogenic environment, e.g., spin qubits in donor silicon and MOS-type double quantum dots.
        To scale up these platforms, controlling qubits in a cryogenic environment is mandatory in the future~\cite{tannu2017taming}.
        Co-integration of qubits and control electronics would circumvent relevant bottlenecks in scalability, such as the cabling and the communication overhead between qubits and their electronic interface~\cite{Sebastiano17}.
        Unfortunately, co-locating the electronics may be limited in cryogenic qubit platforms, where for typical dilution refrigerators the cooling power is $\mathcal{O}(1)$~mW below 100~mK and it is several watts at 1-4~K.
        In a system with thousands of qubits, even the latter leaves at most a few~mW of cooling power per qubit.
        Note that placing qubits and electronics/decoding at two different temperature stages, for example at 20~mK and 4~K respectively, does not really solve the scaling issue related to cabling since the main bottleneck remains across those two stages.
        
        In terms of hardware costs for decoding, power consumption and die area are the most relevant metrics (see~\cref{tab:cryo_decoder_comparison}), especially when operating the decoder very close to the quantum processor, i.e., at the same operating temperature or even on the same die or package.
        We discuss challenges and opportunities for CMOS and SFQ.
        
        \textbf{Cryo-CMOS.}
        Cryogenic CMOS (cryo-CMOS) technologies for peripherals of quantum computers in a cryostat have been actively studied recently~\cite{Patra20,Park21,Bardin19}.
        For example, Ref.~\cite{Patra20} proposed a cryo-CMOS qubit controller fabricated in Intel 22-nm FinFET technology, and its power consumption is several hundred mW.
        However, as superconducting transmon qubits operate below 100~mK, dilution refrigerators offer insufficient cooling power for these electronic circuits at that temperature (as discussed above), although recent efforts propose an electronic interface for transmons operating at 4~K~\cite{Frank22,Kang22}.
        Alternatively, semiconductor spin qubits can operate at temperatures above 1~K~\cite{Petit18}, where the power budget is enough for SoCs~\cite{Patra20,Park21,Bardin19}, making them in principle compatible with standard CMOS~processing.

        \textbf{SFQ.}
        Although an SFQ~decoder can in principle operate at the millikelvin stage, all of the SFQ~decoders proposed so far are assumed to operate in a 4~K environment since the existing SFQ~process technologies are designed for that. 
        In particular, these technologies do not allow to meet the required power consumption at the millikelvin layer, as even the lowest-power SFQ-based decoder, namely the Clique decoder~\cite{ravi2022better}, uses 99~$\mu$W for a distance-9 logical qubit (see~\cref{tab:cryo_decoder_comparison}).
        While satisfactory in the near-term, this is insufficient in a large FTQC~architecture with thousands of logical qubits.
        To realize such architecture using SFQ~circuits, it is thus essential to reduce their power consumption.
    
        One solution is a new SFQ~process technology targeting a millikelvin environment.
        If targeting such environments, it is theoretically possible to reduce the switching energy of an SFQ~circuit by two orders of magnitude compared to the 4~K~environment case.
        However, careful parameter selection according to the fabrication size of Josephson junctions and to the target operating frequency is required. 
        Recently, the adiabatic quantum flux parametron~(AQFP)~\cite{takeuchi2013adiabatic} and the reversible QFP~(RQFP)~\cite{takeuchi2014reversible} using~AQFP have been proposed as ultra-low-power SFQ~circuits.
        The RFQP~logic achieves its low power consumption by using logically and physically reversible logic gates so that the entropy of information does not change during operation.
        Since the switching energy of~AQFP and~RFQP is more than two orders of magnitude less than that of typical SFQ~circuits, they are expected to be helpful in building a large-scale FTQC~architecture in the millikelvin layer.

            \begin{table}[t]
            \centering
            \caption{Comparison of Cryo-CMOS and SFQ decoders.
            The area, power consumption and throughput are per distance-9 logical qubit. \label{tab:cryo_decoder_comparison}}
            \begin{tabular}{|c|c|c|c|c|c|c|} \hline
            &\begin{tabular}[c]{@{}c@{}} NN\\\cite{Overwater22}\end{tabular}
            &\begin{tabular}[c]{@{}c@{}} AQEC\\\cite{Holmes20}\end{tabular}
            &\begin{tabular}[c]{@{}c@{}} QECOOL\\\cite{Ueno21}\end{tabular}
            &\begin{tabular}[c]{@{}c@{}} QULATIS\\\cite{ueno2022qulatis}\end{tabular} 
            &\begin{tabular}[c]{@{}c@{}} NEO-QEC\\\cite{Ueno22}\end{tabular} 
            &\begin{tabular}[c]{@{}c@{}} Clique\\\cite{ravi2022better}\end{tabular} 
            \\ \hline \hline

            \begin{tabular}[c]{@{}c@{}} Platform\end{tabular}
            & CMOS
            & SFQ
            & SFQ
            & SFQ
            & SFQ
            & SFQ
            \\ \hline
            
            \begin{tabular}[c]{@{}c@{}} Meas.\\errors\end{tabular}
            &
            &
            &\checkmark  
            &\checkmark
            &\checkmark
            &\checkmark
            \\ \hline

            \begin{tabular}[c]{@{}c@{}} Lattice\\surgery\end{tabular}
            &
            & 
            &
            & \checkmark 
            & \checkmark 
            & 
            \\ \hline

            \begin{tabular}[c]{@{}c@{}} Area\\($\text{mm}^2$) \end{tabular}
            & 10
            & 369
            & 183
            & 16.4
            & N/A
            & 14.4
            \\ \hline

            \begin{tabular}[c]{@{}c@{}} Power\\consumpt.\\($\mu$W)\end{tabular}  
            & 20000
            & 3780
            & 400.3
            & 417.4 
            & 614.9
            & 99
            \\ \hline
            
            \begin{tabular}[c]{@{}c@{}} Throughput\\Max/Avg.\\(ns)\end{tabular} 
            & 28
            & 19.2/3.8
            & 364/9.15
            & 82/2.12
            & N/A
            & 0.24
            \\ \hline

            \end{tabular}
            \end{table}

    \section{Outlook}
    \label{sec:conclusion}
        
        Many quantum-computing platforms are now getting to the size and fidelities required for QEC~demonstrations.
        While a number of challenges lay ahead, the dream of FTQC is closer than ever.
        We expect to first see demonstrations for a single logical qubit, likely in FPGA and with sliding-window decoding, and then for logical two-qubit operations.
        The earliest demonstrations will still make use of look-up table decoders, but these will be quickly supplanted by scalable algorithmic decoding.
        The initial code distance will be~3 but will soon scale to~5 and beyond depending on experimental progress.
        By 2025 we expect exciting experimental and theoretical results that address many of these challenges: the development of new decoding algorithms for topological, color and LDPC~codes, improved suppression of realistic noise, demonstrations of decoders tightly integrated with control systems, and real-time decoding of logical operations.
        
        We hope that by 2025 the community will have undeniably demonstrated that utility-scale fault-tolerant quantum computation is achievable in practice, and a clear path will be paved to realize such systems in the second half of the decade. All of this will likely be demonstrated on a variety of qubit technologies, each of which is likely to face different challenges throughout the stack.
        For real-time decoding this means that different solutions will be explored, from running on CPUs to tighter integration on FPGAs or ASICs.
        Time will tell if FPGA~solutions can offer scalability and speed comparable with ASIC~solutions, eventually co-integrated with qubit control and readout electronics.
        Following the current trend towards more compact large-scale computing systems, the electronics, including the decoder, will eventually operate close to the qubits (at cryogenic temperature for the solid-state qubit platforms).
        We may also see some experiments performed with ASIC-based decoders within a cryogenic environment.
        While ASICs may not be essential to the experimental success as of 2025 yet, they could be used to showcase the benefits of reducing the overall power consumption of the system and pave the way forward for the fault-tolerant era of highly integrated, scalable and cost-effective quantum systems.
        
\section{Acknowledgements}
    \label{sec:acknowledgement}
    Qblox is supported by the European Commission, Grant agreement ID: 969201.
    Y.U.~is supported by Grant-in-Aid for JSPS Research Fellow Grant Number JP21J10882.

    
\providecommand{\noopsort}[1]{}\providecommand{\singleletter}[1]{#1}%

\end{document}